\documentclass[prl,twocolumn,superscriptaddress,showpacs,floatfix]{revtex4-1}
\usepackage{graphicx}
\usepackage{dcolumn}
\usepackage{bm}
\usepackage{epsfig}
\usepackage{wasysym}
\usepackage{color}
\usepackage{hyperref}

\begin{document}

\title{Understanding and modulating the competitive
    surface-adsorption of proteins}
\date{\today}

\author{Pol Vilaseca} 
\affiliation{Departament de F\'{\i}sica Fonamental, Facultat de F\'{\i}sica,
Universitat de Barcelona, Diagonal 645, 08028, Barcelona, Spain}
\affiliation{School of Mathematics, Trinity College Dublin, Ireland}

\author{Kenneth A. Dawson}
\affiliation{Center for BioNano Interactions (CBNI), University College of Dublin, Ireland}
\email{Kenneth.A.Dawson@cbni.ucd.ie, gfranzese@ub.edu}

\author{Giancarlo Franzese}
\affiliation{Departament de F\'{\i}sica Fonamental, Facultat de F\'{\i}sica,
Universitat de Barcelona, Diagonal 645, 08028, Barcelona, Spain}


\begin{abstract}

It is now well accepted that 
cellular responses to materials in a biological medium reflect
greatly the adsorbed biomolecular layer, rather than the material
itself.
Here, we study by molecular dynamic simulations the competitive protein
adsorption on a surface (Vroman effect), i.e. the non-monotonic behavior of the amount of
protein adsorbed on a surface in contact with plasma as a function of 
 contact time and plasma concentration. 
We find a complex behavior, with regimes during which small and large
proteins are not necessarily competing between them, but are both
competing with others in solution. 
We show how the effect can be understood, controlled and inverted.
\pacs{87.10.Tf, 
81.16.Fg, 
87.85.J-
}

\end{abstract}

\maketitle

\section{Introduction}


\footnotetext{\textit{$^{a}$~Departament de F\'{\i}sica Fonamental, Facultat de F\'{\i}sica,
Universitat de Barcelona, Diagonal 645, 08028, Barcelona, Spain.
E-mail: gfranzese@ub.edu}}
\footnotetext{\textit{$^{b}$~Center for BioNano Interactions (CBNI),
    University College of Dublin, Ireland.
E-mail: Kenneth.A.Dawson@cbni.ucd.ie
}}




When nanoparticles are in contact with blood plasma, or other biological fluids,
biomolecules rapidly coat the  bare surface in a relatively selective
manner \cite{Ced2007}. It is increasingly accepted that the early
biological responses to nanoparticles will be determined by the  
adsorbed biomolecules rather than the pristine surface alone \cite{Mono2011,Kim2012,Wal2010}.
Because of their size \cite{Mono2011,Ober2009}
nanoparticles are trafficked by active transport processes throughout 
the organism, using the information from the protein sequences
associated with the surface of  nanoparticles.
Unlike the situation for flat macroscopic surfaces say of medical implants, for nanoparticles
the protein environment changes in different compartments of cells and organs, as
the nanoparticle traffics. This has lent urgency to the modern
interest in understanding the phenomenon at a more fundamental level
\cite{Mono2011}. Still, we can learn a lot from an understanding of
the process for flat surfaces \cite{Tengvall1998}.
Studying the adsorption of Fibrinogen on a surface in contact with
blood plasma, Vroman found that the surface concentration of
Fibrinogen shows a maximum at an intermediate contact time, indicating
that Fibrinogen is replaced with time by one or more families of
different proteins \cite{VROMAN1962}.  
The phenomenon is not specific to Fibrinogen, but is a general effect
for many other proteins \cite{Ortega1995,Holmberg2009}.  The plasma proteins 
compete for the occupation of the surface, resulting in a 
sequential competitive adsorption, known as the Vroman effect. 

The effects depends on numerous factors such as the 
plasma dilution, the pH, the temperature, the surface charge and the
specific surface chemistry \cite{Jackson2000}. 
In highly concentrated plasma, the sequential adsorption takes place
in seconds, but  it takes several minutes when the plasma is diluted
\cite{LeDuc1995}.
The effect has been documented both on hydrophilic and 
hydrophobic interfaces \cite{LeDuc1995,Green1999}
being more evident the more hydrophilic the material, but with
stronger protein binding the more hydrophobic the surfaces
\cite{Noh2007,Ding2011}. However, no universality is found in the
experiments and the results strongly depend on the 
details of the experiments \cite{Lassen1996,Lassen1997,Barnthip2009}.
It is generally accepted that proteins with smaller molecular
weight and at higher concentration adsorb first to the surface, but
later are replaced by other proteins with, generally, larger molecular
weight and size. After the adsorption, the protein can undergo
conformational changes and denaturation, especially at a hydrophobic
interface, eventually leading to irreversible adsorption
\cite{Green1999}.

Many experimental techniques have been used to investigate the effect
from blood plasma or model solutions with a limited number of
components and many models have been proposed to rationalize the
experiments
\cite{Lu1994,Brash1995,Slack1995,LeDuc1995,Dejardin1995,Green1999}. 
However, the mechanisms of the phenomenon are still debated and no
existing model can fully explain it 
\cite{Jung2003,Barnthip2009,Hirsh2013}. 

Volumetric effects, due to non-deformable proteins trying to fit on the available
surface, can account for competitive adsorption of proteins 
\cite{Barnthip2009}. However, they do not reproduce the maxima of
absorption of the Vroman effect. This maxima is, instead, rationalized
by models based on kinetic equations. Some of these models include coupled
mass transport equations \cite{Lu1994}. In all of them, to each kind of
protein in solutions, there are associated  
different adsorptions/desorption rate constants. These processes are
modeled as reversible by some authors \cite{CUYPERS1987}. Others, to fit
better the experiments, assume that the adsorption  can become irreversible 
with a ``reaction'' rate constant \cite{Lu1994}. Due to the difficulty
for this approach to describe the variety of experimental results,
some models include also a ``displacement'' rate constant of a reversibly
adsorbed protein by a protein with a higher surface affinity
\cite{Lundstrom1990,Slack1995,Dejardin1995}. However, 
these models are unable, in general,  to describe solutions at low
concentration, where the 
surface coverage is controlled by diffusion \cite{Dejardin1995}, and
cannot  rationalize the different desorption behavior observed
for sorbent-free with respect to sorbent-bearing washing
solutions \cite{LeDuc1995}. 

The latter observation inspired LeDuc et al. to include also a
``liberation'' rate constant of semipermanently adsorbed protein by
contact with a bulk protein \cite{LeDuc1995}. To simplify the model,
the authors made strong approximations, likely to be incorrect,
assuming that adsorbed proteins do not diffuse on the surface and that
the displacement and liberation rate constant do not depend on the
incoming protein \cite{LeDuc1995}. They applied the model to
rationalize data of a ternary solution mixture with Albumin, high
molecular weight Kininogen and Fibrinogen, accounting also for the
deformation of the semipermanently adsorbed proteins. As a result,
LeDuc et al.  found that, to fit the data, the first two proteins
should occupy approximately fourfold more space in the semipermanent
state while Fibrinogen would have a much smaller change.

This is at variance with what recent experiments show for rod-like
proteins as the Fibrinogen. This elongated protein, although deforms
less then Albumin when adsorbed on an extended surface, can undergo a
large rearrangement from an initial ``lying down'' stage (with its
long axis parallel to the surface) to a ``standing up'' conformation
(long axis perpendicular to the surface). This conformational change
results in a large difference in the occupied surface \cite{Roach2005}.


While the models based on kinetic equations are useful to
qualitatively reproduce the experimental data by fitting the rate
constants, they are less instructive about the mechanisms that at
molecular level control the phenomenon.  To give an insight
into how the competition between sizes, bulk concentrations,
surface affinities, diffusion constants 
and conformational changes combines to give rise to the Vroman effect,
we devise here a coarse-grained model of a ternary protein solution
mixture in contact with a hydrophobic surface.

\section{The Model}

A full atom simulation of competitive adsorption of proteins from a
multicomponent  mixture is at present time unfeasible for several
reasons. Each protein is made of a large number of amino acids (e.g.,
585 for 
human serum Albumin 
and more than 2800 
for the 
human Fibrinogen) and is hydrated by thousands of water molecules. As
a consequence, a fully atomistic Molecular Dynamics (MD) simulations of one
single protein adsorption on a surface with explicit water is limited
to a few hundreds of ns \cite{Wei2012}. This time scale is at least
five orders of magnitude smaller than the one necessary to observe competitive
adsorption. Moreover, the simulations should be for thousands of proteins.

This challenging task can be undertaken by coarse-graining the
system. Coarse-graining can be performed at different levels
\cite{Tozzini2005}. However, modeling a full layer of adsorbed proteins
on an extended surface urges to reduce drastically the degrees of
freedom. A common strategy is to consider implicit water and to
represent the protein as a
single particle. As we will explain in the following, this approach
does not prevent us from taking into account the possibility of
conformational changes. We now describe the details of the model with 
the approximations we make to reduce the complexity of
the problem, bearing in mind that our aim  
is to show that the competitive adsorption can be understood in terms of a
general mechanism, regardless the specific details of the
real interactions in the system.

We consider the three  most abundant proteins in human blood: Albumin
(Alb), Immunoglobulin-$\gamma$ (IgG) and Fibrinogen (Fib), for which  
competitive adsorption on hydrophobic surfaces has been observed
\cite{Bale1988,Lassen1997}. 
The model assumes an implicit solvent and 
includes through effective potentials the specific energetic and 
entropic effects of the water hydrating the proteins 
and the surface \cite{fbi2011}, as well as those effects due to the
charge distribution on the protein surface 
or the counter ions in the solution \cite{Gallo2011}.
This method has been
validated in many specific cases (e.g., see 
\cite{Lund2008,Ravichandran2000}) and follows a general approach that
has led to the well established DLVO theory (e.g., see 
\cite{Oberholzer1997,Ruggiero1999,Bellion2008}).

Alb is a globular protein, with an almost spherical shape and an
 isoelectric point (IEP)  at approximately pH
5.0 \cite{Malmsten1997}. 
By considering a pH 5.0, we minimize the
charge interaction for Alb. Hence, the interaction of Alb with the
surface is modeled with a short range attraction, that can be thought
as mainly due to the entropic gain for water exclusion at the
interface, 
\begin{equation}
 V_{A,S}(z)\equiv
 4\epsilon_{A,S}\left(\left(\frac{\sigma_{A}}{z}\right)^{24}-\left(\frac{\sigma_{A}}{z}\right)^{12}\right)
\label{1}
\end{equation}
where $z$ is the distance between the center of mass of the protein
and the surface, $\epsilon_{A,S}$, related to the binding affinity and
the dissociation constant,  is the attractive energy between 
Alb and the surface. Here $\sigma_{A}\equiv R_A/2^{1/12}$, with $R_A$ radius of the Alb,  
takes into account that Alb is a globular protein whose conformation
may become distorted on interaction with the surface \cite{Roach2005}.

The IgG structure resembles a $\gamma$ that can be roughly
approximated with a spherical shape.
We model protein-protein interaction for the two spherical proteins as 
\begin{equation}
 V_{i,j}(r)\equiv\epsilon_{i,j}\left(\frac{\sigma_{i,j}}{r}\right)^{24}
\end{equation}
where $r$ is the protein-protein distance, $\epsilon_{i,j}$ the
characteristic interaction energy 
between protein $i$ and protein $j$, where each index can be
$A$ for Alb or $I$ for IgG,  and $\sigma_{i,j}\equiv R_i+R_j$,
with $R_i$ radius of protein $i$.
Attraction among proteins is not included at this level of description,
as it is small compared to protein-surface interaction and the protein
solution is stable \cite{Lu1994}. 

\begin{figure}
\centering
\includegraphics[clip=true,scale=0.25]{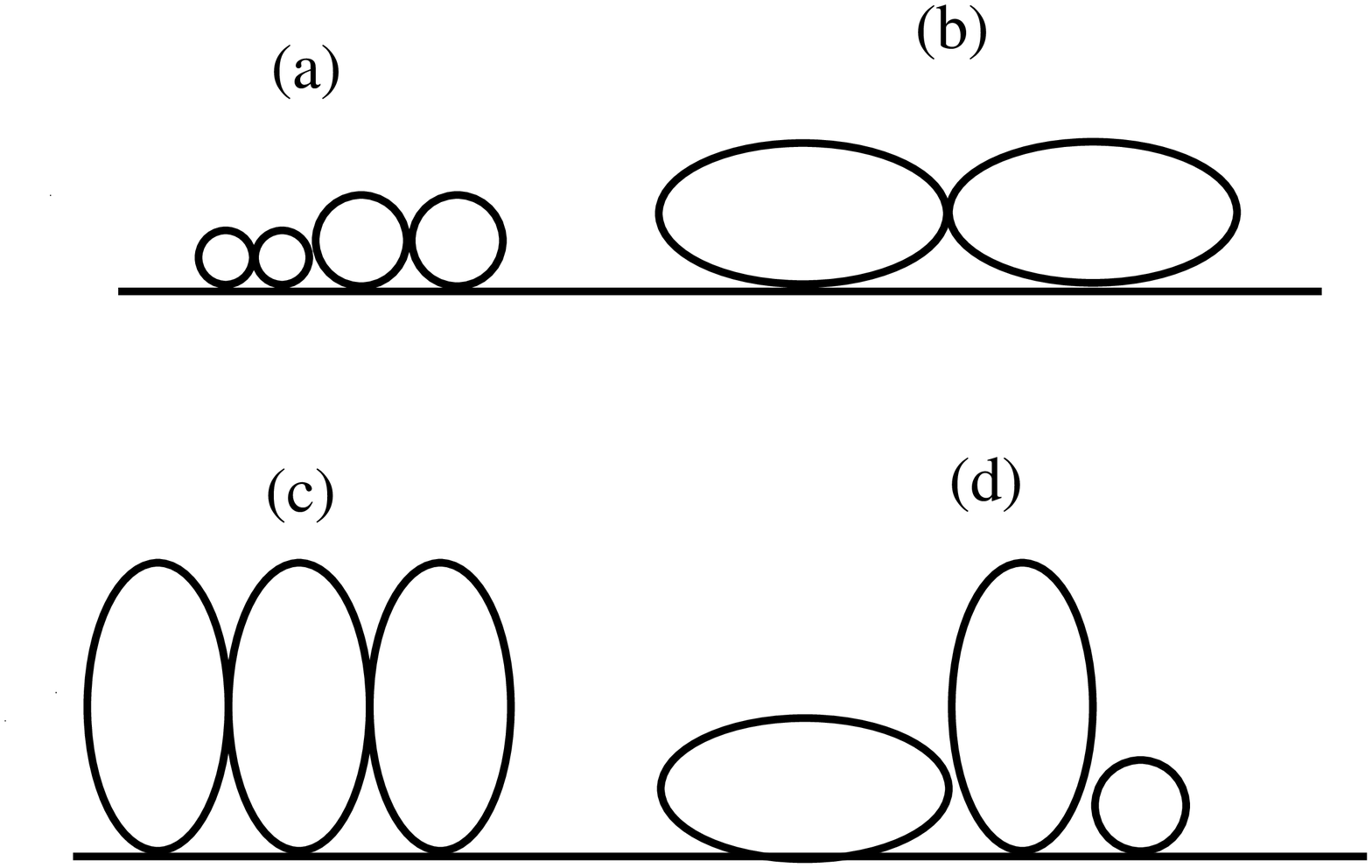}
\caption{Schematic representations of different proteins adsorbed
  on the surface. 
 (a) Alb (smaller) and IgG (larger) are approximated as globular
 proteins with radiuses $R_A<R_I$.
In all the panels the continuous line represents the 
surface profile. 
(b) Fib is represented as an
ellipsoid with a short axis 
$R_F$ and a long axis $\delta_F$. Fib can assume different
conformations: ``lying down'', as in (b), or  
``standing up'', as in (c), possibly giving rise to mixed configurations, as in (d).} 
\label{Fig1}
\end{figure}

The Fib in its folded conformation is rod-like. We approximate it with 
an elongated ellipsoid,  with two 
principal axes of rotation, that can assume two different
conformations, one  ``lying down'' and another ``standing up'' on the
surface (Fig.~\ref{Fig1}). This idea is consistent with experiments
\cite{Roach2005} and has been used in Monte Carlo simulations with 
potentials within the DLVO theory \cite{Bellion2008}.
Here the two  different conformations are
encoded in an effective way through soft-core Fib-``protein $i$'' potentials, 
\begin{equation}
 V_{F,i}(r)\equiv\epsilon_{F,i}\left[\left(\frac{\sigma_{F,i}}{r}\right)^{24}+\frac{3}{1+\exp\left(30\left(r-\delta_{F,i}\right)/\sigma_{A}\right)} \right]
\label{2}
\end{equation}
where 
$i=A, I, F$ stands for Alb, IgG and Fib, 
with $\sigma_{F,i}\equiv R_F+R_i$ corresponding to
the interaction along the short axis,
$\delta_{F,i}\equiv\delta_F+\delta_i$
corresponding to the interaction along the long axis,
$\delta_A\equiv R_A$, 
$\delta_I\equiv R_I$, 
$\delta_F$ the long axis of Fib,  
and $\epsilon_{F,i}$ the characteristic interaction energy of Fib with
the protein $i$.
The protein-protein interaction with the Fib along the short axis is
chosen 
energetically unfavorable with respect to that along the long axis,
because the latter offers more binding points to the surface.

Since both (monoclonal) IgG \cite{Galisteo1994} and Fib \cite{Malmsten1997} have an
IEP at approximately pH 5.5, at the chosen pH 5.0 they are charged.
Following other authors, e.g., Ref.~\cite{Mucksch2011},  
we consider that the charged proteins, IgG and Fib, have an effective
interaction with the surface modelled by a Lennard-Jones potential
\begin{equation}
 V_{i,S}(z)\equiv
 4\epsilon_{i,S}\left(\left(\frac{\sigma_{i}}{z}\right)^{12}-\left(\frac{\sigma_{i}}{z}\right)^{6}\right) 
\end{equation}
where $\epsilon_{i,S}$ is the attractive energy between 
the protein $i=I, F$ and the surface, and $\sigma_{i}\equiv R_i/2^{1/6}$ accounts for
the possible distortion of the protein in contact with the surface. 

When adsorbed, each protein occupies a surface $2\pi R_i^2$, including
the ``standing up'' conformation of Fib, while the Fib in its ``lying
down'' conformation occupies a surface  $2\pi \delta_F^2$.


To account for the different diffusive behaviors of different proteins
in absence of external flow, we calculate the hydrodynamic radius
$R^{H}_i$  of each protein $i$, under the
assumption that the proteins can be approximated by a sphere,
 through the Einstein-Stokes
equation $D_i=\frac{k_{B}T}{6\pi\eta  R^{H}_{i}}$,  where $D_i$
is the experimental  diffusion coefficient, $\eta$ the viscosity of
the medium, $k_{B}$ the Boltzmann constant, $T$ the absolute temperature.
Next, we identify $R_A=R^{H}_A$, $R_I=R^{H}_I$, and
$\delta_F=R^{H}_F$, while $R_F$ is
set by imposing  that the experimental surface
concentration found for Fib corresponds to its close packing
configuration in the ``standing up'' conformation \cite{Lu1994}.
These conditions give 
$R_{A}=3.55$~nm, $R_I=5.51$~nm,  $R_F=9.29$~nm 
and $\delta_F=11$~nm.
Protein masses 
$M_{A}=67$~kDa, 
$M_{I}=150$~kDa, 
$M_{F}=340$~kDa, 
necessary to determine the time scales, are known from
experiments \cite{Noh2007}. 

Because we include only repulsive protein-protein interaction, for
sake of simplicity  we set all the protein-protein
$\epsilon_{i,j}=\epsilon_{A,S}$. 
Protein-surface attraction energy
$\epsilon_{i,S}$ can be calculated 
from the adsorption rate constants \cite{Lu1994}.
These rates 
are proportional to the probability for a  
 protein $i$ to attach to the nearby surface
\begin{equation}
 P_{i}\propto \exp\left(\frac{\epsilon_{i,S}}{k_{B}T}\right).
\end{equation}
However, the $\epsilon_{i,S}$ 
in physical units  are not known {\it a
priori}. Hence, we consider the relative probabilities for different proteins
$\frac{P_{i}}{P_{j}}\propto
\exp\left(\frac{\epsilon_{i,S}-\epsilon_{j,S}}{k_{B}T}\right)$, from which
is possible to determine the values of the different energies as 
\begin{equation}
 \frac{\epsilon_{j,S}}{\epsilon_{A,S}}=1-
\frac{k_{B}T}{\epsilon_{A,S}}\ln\left(\frac{P_{A}}{P_{j}}\right)
\label{5}
\end{equation}
adopting $\epsilon_{A,S}$ for Alb as the energy units. We set
$\epsilon_{A,S}$, the only free parameter of our model, 
by comparing our simulations results with experiments 
at ambient temperature, and get 
$\epsilon_{I,S}=2.79~\epsilon_{A,S}$ 
and $\epsilon_{F,S}=6.08~\epsilon_{A,S}$ 
by adopting the 
adsorption rate constants as in the theoretical model of Lu et
al. \cite{Lu1994}, that reflect the experimental observation that Fib
has the strongest affinity for several surfaces and albumin the
weakest.  

\section{The Numerical Method}

We perform MD simulations at constant $T$, constant volume $V$ and
constant number of proteins $N_i$, in a
parallelepiped with two square faces and four 
rectangular faces. A square face is occupied by the attractive
surface, the other 
by a wall interacting  with the proteins through the repulsive part of
the potential in Eq.(\ref{1}). We apply  periodic boundary conditions
(pbc) along the four 
rectangular faces.
The volume concentrations of proteins is taken to match the average
concentrations of the human plasma, with
$c_{A}=4.25$~g/dl, 
$c_{I}=1.25$~g/dl 
and $c_{F}=0.325$~g/dl,
at $X_P=100\%$ plasma concentration in blood.
When a protein is adsorbed on (released by) the surface, we keep its bulk
concentrations constant by inserting (deleting) a protein of the same
family in a randomly-chosen empty (occupied) space of the box.

Experiments are usually carried out for highly diluted plasma, 
at concentration as small as $X_P=0.1\%$, 
to slow down the adsorption rate 
to minutes or hours, 
allowing 
precise measurements.
However, 
such low rates would decrease the statistics of our 
MD simulations.
We, therefore, perform our simulations in conditions that are closer
to those of practical interest, with $X_P$ as high as 
$100\%$, $50\%$ and $25\%$, by considering different sizes of the simulation box
while keeping constant the initial number of proteins, their  relative
proportions, and the size of the adsorption surface. 
For each $X_P$ 
we average the results over 
fourteen independent runs, starting from independent 
initial configurations that have been equilibrated by applying pbc in any direction.

\section{Results and Discussion}

\subsection{Competitive Adsorption}

\begin{figure}
\centering
\includegraphics[clip=true,scale=0.5]{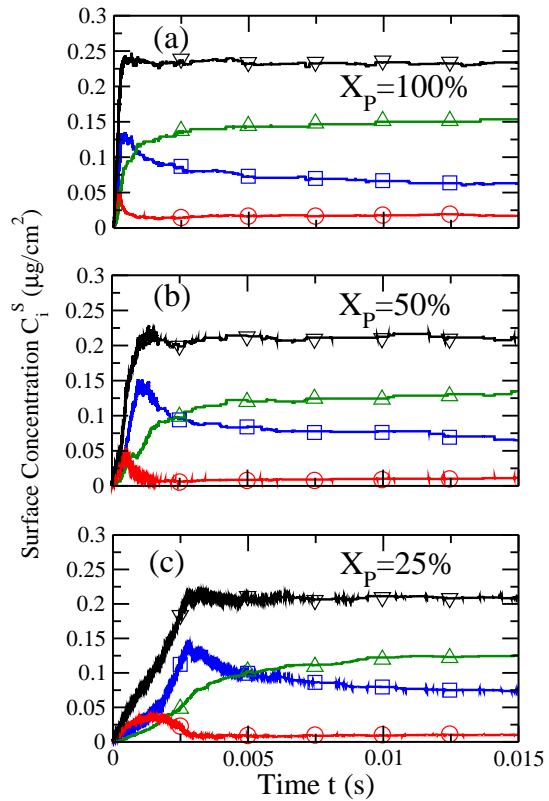}
\caption{Simulations at $T=300$~K and (a) $X_P=100\%$, (b) $X_P=50\%$
and (c) $X_P=25\%$ show that, at any
considered dilution, surface concentration $C^S_A$ of Alb
(\Circle), $C^S_I$ of IgG (\Square) and $C^S_F$ of Fib ($\Delta$) are
not all monotonic with time, while their sum ($\nabla$) is monotonic
within our numerical precision. 
For sake of clarity we plot the symbols only for a limited number of times.
Bulk concentrations are as indicated in the text.
Errors are smaller than symbol sizes. }
\label{f1}
\end{figure}

We find that protein surface concentrations $C^S_i$, for Alb and IgG,
are non-monotonic in time (Fig.~\ref{f1}). In particular,
for any considered $X_P$,  Alb is the first protein that reaches 
the surface, due to its larger diffusive constant. This property
induces an increase of  $C^S_A$. When the second fastest
and second most affine protein, IgG, diffuses to
the surface, it displaces Alb, leading to a decrease of $C^S_A$
and an increase of $C^S_I$. Finally 
Fib,  which is the slowest and most affine protein to the
surface, takes over decreasing $C^S_I$ and increasing $C^S_F$.
Each $C^S_i$ saturates toward  an equilibrium value
at long times, while the total surface concentration of proteins is
saturated at early times. This behavior qualitatively reproduces the
Vroman effect, apart from the behavior of Fib that here is monotonic, while in
experiments has a maximum due to the competitive adsorption with heavier
and more surface-affine plasma proteins, like the high molecular
weight Kininogen, not included in our model 
\cite{LeDuc1995,Turbill96}. 

\subsection{Effect of Plasma Dilution}

\begin{figure}
\centering
\includegraphics[clip=true,scale=0.5]{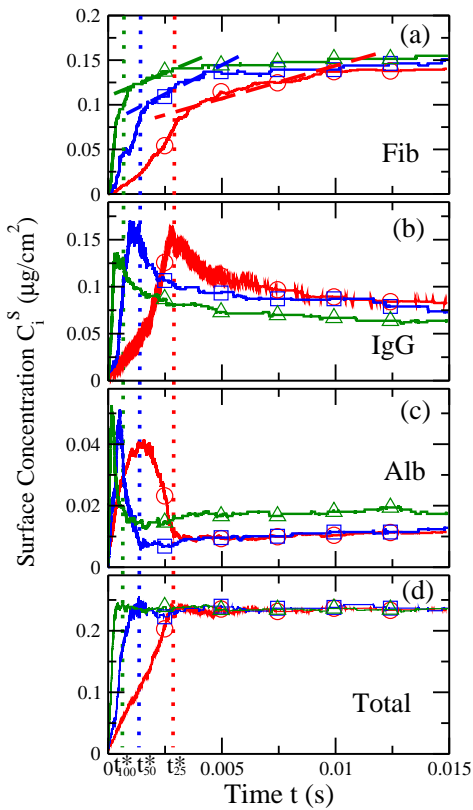}
\caption{Same data at  $T=300$~K  as in in Fig.~\ref{f1} 
now separated for (a)  Fib, (b) IgG, (c) Alb, (d) and their sum
at $X_P=100\%$ ($\Delta$),
  $X_P=50\%$ (\Square), and $X_P=25\%$ (\Circle). 
The time-scales increase for  decreasing $X_P$. 
In (a) the dashed lines are guide to the eyes for the linear regime of
the three-steps kinetics of Fib adsorption. The vertical dotted lines
are guides for the eyes to mark characteristic times.
Note that the vertical scale of panel (c) is almost four times smaller
than those in panels (a) and (b).
} 
\label{fdil300}
\end{figure}

When $X_P$ is reduced the dynamics of the process
slows down (Fig.~\ref{fdil300}).  This is consistent with what 
is observed in experiments \cite{Brash1984}  and was reproduced 
by kinetic models with displacement rates, ``liberation''
rates and semipermanently adsorbed state for Fib  \cite{LeDuc1995}. Here we
can observe the slowing down not only for Fib, but also for the
competing proteins IgG and Alb. 

It is interesting to observe that at short-times the surface
concentration of Fib increases more than linearly with time. This is
more evident at low $X_P$   (Fig.~\ref{fdil300}a). This behavior has
been predicted in other models for single protein adsorption including
conformational changes and has been noted that it
is not reproduced by standard Langmuir kinetics
\cite{Adamczyk1995}. It can be understood as a consequence of the
ability of Fib to adsorb in both its ``laying down'' and ``standing
up'' conformation that is not captured by standard Langmuir
kinetics. This stage can be considered as the first step of Fib
adsorption and occurs when the competitions with the other proteins is
not strong, i.e. when the total surface concentration has not reached
its saturation (Fig.~\ref{fdil300}d).

\subsection{Three-steps adsorption}

By increasing the dilution
(i.e. reducing $X_P$) all the surface concentrations 
tends toward the same large-time limit (Fig.~\ref{fdil300}). 
Alb concentration reaches
a shallow minimum at a time $t^*_{X_P}$ that depends on $X_P$ and
approximately corresponds to that 
of maximum surface concentration of IgG. 

This time $t^*_{X_P}$
coincides, within our numerical precision, also with the beginning of
a ``linear'' regime, i.e. a regime of constant adsorption rate, for
the Fib. This linear regime represent a second step in the Fib
adsorption and precedes a third step during which $C^S_F$ saturates.
This three-steps
kinetics has been experimentally observed, and
numerically reproduced, in single protein adsorptions on thin SiO$_2$ layers, 
both hydrophilic or with an additional hydrophobic monolayer, 
at room temperature and at 37.5 $^\circ$C \cite{Quinn2008,Bellion2008}.  
The authors of those works interpreted this behavior
as a consequence of protein diffusion at the surface and of
the occurrence of  conformational changes. However, they did not study
the case with competitive adsorption. 

Here, instead, we observe that
the regime of constant Fib adsorption rate coincides  with the IgG
desorption and the slow re-adsorption of Alb. This suggests that the
reorganization of the proteins at the surface is likely to involve all
the three families of proteins at the same time, in a way that is
far more complicated than the usual two-states models 
based on kinetics equations with ``transition'' rate constants. 
Indeed, standard Langmuir kinetics would be able to predict the general trend
of slowing down for increasing dilution \cite{Masel}, but 
is unable to reproduce the three-steps kinetics, even in single protein
adsorption, when  conformational changes take place
\cite{Adamczyk1995,Quinn2008,Bellion2008}.

The second step starts, at $t^*_{X_P}$, 
when the total surface concentration is saturated (Fig.~\ref{fdil300}d). Therefore, new
arriving Fib adsorbs in the ``laying down'' conformation if possible, 
or, with less energy gain but occupying less space, in the ``standing up''
conformation. Since at $t^*_{X_P}$ the IgG concentration is at its
maximum, the probability that the new Fib adsorbs near a IgG (as in
Fig.\ref{Fig1}d) is high, determining a strong repulsion between
the two charged proteins.  This repulsion is stronger than the
attraction of IgG with the surface, determining the displacement of
the IgG and the decrease of $C^S_I$. 

This displacement leaves enough space on the surface for the
adsorption of the smaller Alb that is abundant in suspension. As a
consequence, $C^S_A$ increases. Despite Alb lower affinity to the
surface, its small size allows the protein to fit onto the free
surface without experiencing strong repulsion with the Fib. Therefore,
at this stage Alb and Fib are not necessarily competing between them,
but are both competing with the IgG. 

However, at larger time, when
more Fib arrives to the surface, the competition is strong among all
the three proteins. This induces  the end of the re-adsorption of
Alb and forces further conformational changes for the Fib (as in
Fig.\ref{Fig1}c). Our calculations support the identification of
the third adsorption step, i.e. the end of the regime of constant
Fib adsorption rate, with the end of the re-adsorption of Alb. This is
more evident for the the lowest dilution, $X_P=100\%$, while is more
speculative for the other values of $X_P$.


\begin{figure}
\centering
\includegraphics[clip=true,scale=0.5]{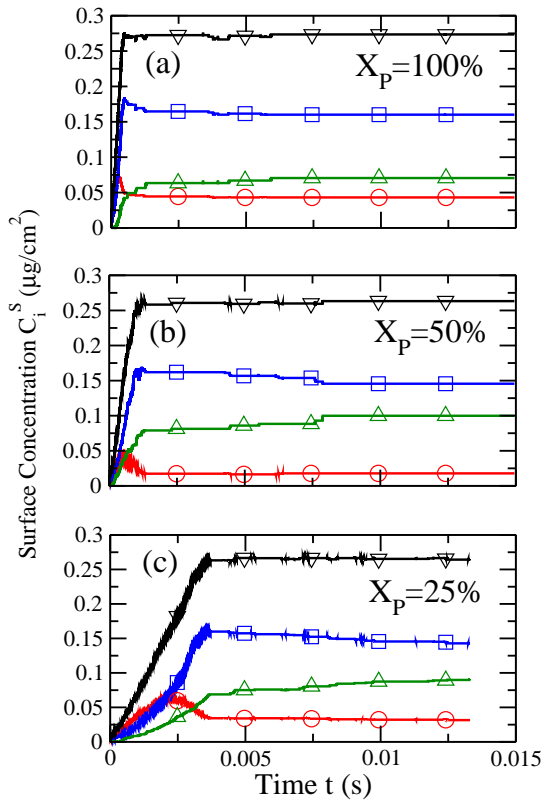}
\caption{Surface concentrations $C^S_i$ as function of time for
  $T=120$~K at (a) $X_P=100\%$, (b) $X_P=50\%$  and (c) $X_P=25\%$. 
At long time,  $C^S_I>C^S_F$, with an inversion with respect to
the standard conditions in Fig.~\ref{f1} where $C^S_F>C^S_I$.
Errors and symbols are as in Fig.~\ref{f1}:
Alb (\Circle), IgG (\Square), Fib ($\Delta$) and
their sum ($\nabla$).
} 
\label{f2}
\end{figure}

\subsection{Effect of Energy Depletion}

Next, we study how energy depletion of the protein solution
affects the sequence of adsorption.
In experiments the energy is controlled by adding sodium azide, or other 
depletion-energy
chemical agents, to the protein solution \cite{Shapero2011}. 
Here, for sake of simplicity, we
decrease $T$, reducing the kinetic energy of the solution, but
neglecting possible effects of the protein stability (Fig.~\ref{f2}).

We find ($i$) that, although the surface affinity of
Fib is stronger than that for IgG, the
latter becomes the dominant protein adsorbed on the surface for long
time scales; ($ii$) 
that, by changing $X_P$, the time
scale of the process becomes longer, but the inversion of the protein
concentration is always present.
Hence, 
the energy depletion leads  to an inversion of the Vroman effect. 

By comparing the results at different energies, $k_BT$, and same $X_P$
(Fig.~\ref{f1}-\ref{f2}),
we observe only a  week energy-dependence of the times at which each
$C_i^S$ reaches its maximum. Hence,
the time-scales of the process are mainly
controlled by the total plasma concentration $X_P$, while the
slowing-down due to the reduced diffusion seems to be less relevant.

\begin{figure}
\centering
\includegraphics[clip=true,scale=0.5]{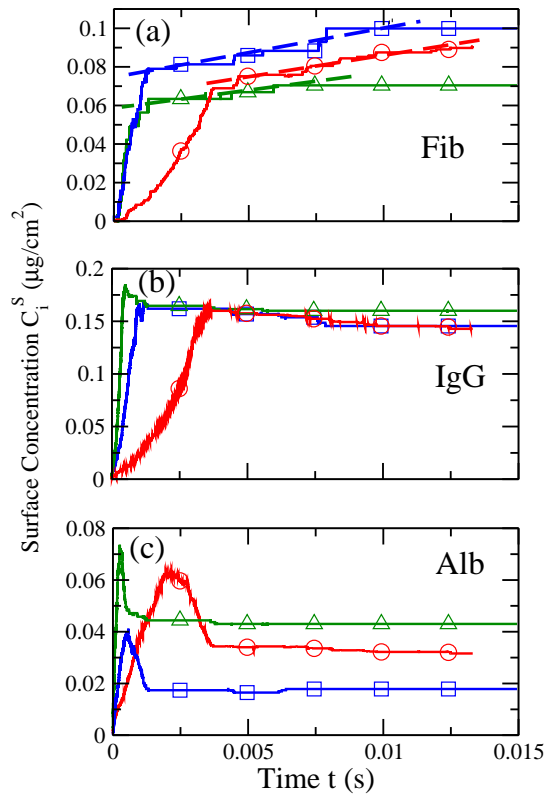}
\caption{Same data at  $T=120$~K as  in Fig.~\ref{f2}
now separated for (a)  Fib, (b) IgG and (c) Alb, at  $X_P=100\%$ ($\Delta$),
  $X_P=50\%$ (\Square), and $X_P=25\%$ (\Circle).
In (a) the dashed lines are guide to the eyes, showing that the 
linear regime is more extended than at $T=300$~K.
For Alb the  $C^S_A$ of saturation is non-monotonic with $X_P$.
} 
\label{fdil120}
\end{figure}

By comparing the $C_i^S$ at different $X_P$ for the same protein
(Fig.~\ref{fdil120}), we find that second step in the Fib adsorption
is now more extended in time. This result is consistent with what has
been observed in the experiments from single protein absorption when
conformational changes occur  \cite{Bellion2008}. Moreover, our analysis
for competitive adsorption shows, as for $T=300$~K, that the
linear regime of Fib adsorption coincides with the end of the
desorption of the Alb and the beginning of desorption of IgG. However,
at $T=120$~K, the ability of Fib to displace IgG is much more limited
than at $T=300$~K, because the displacement requires too much energy.
This fact, on one hand, limits the adsorption of Fib, on the other
hand does not allows Alb to re-adsorb. Nevertheless, the competition
between Fib and IgG is enough to stop the desorption of Alb that now
saturates at a value higher than for $T=300$~K at any $X_P$. 

Another not intuitive result is that at $T=120$~K the adsorption
behavior is less regular than at $T=300$~K. For example, at the lowest
$X_P=25\%$ the Fib seems to adsorb more than at the highest,
$X_P=100\%$, and less than at the intermediate $X_P=50\%$. A similar
non-monotonic behavior characterizes also the Alb absorption, but now
the $C_A^S$ is higher when $C_F^S$ is lower and vice versa. These results suggest
that, under this condition, the strongest competition is between Alb
and Fib, because IgG is almost not displaced from the
surface. Furthermore, at $X_P=25\%$ Fib does not reach the third step 
of adsorption, suggesting that the kinetics is so slow that it
does not allow the Fib to perform large conformational changes.

\subsection{Change of the Bioenvironment}

\begin{figure}
\centering
\includegraphics[clip=true,scale=0.5]{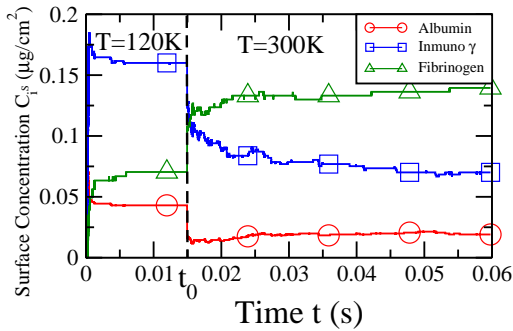}
\caption{The surface concentrations $C_i^S$, as a function of time for
  $X_P=100\%$, is drastically affected when the system undergoes a
  sudden change from en energy-depleted condition to a normal condition.
The vertical dashed line marks the time $t_0$ of the change.
We control the energy of the solution by changing the external
parameter $T$ from $T=120$~K to $T=300$~K.
Errors and symbols are as in Fig.~\ref{f1}.} 
\label{f6}
\end{figure}

Once we have understood that the protein layer covering the surface 
is controlled by the energy depletion of the system, it
is interesting to ask if a sudden change of external conditions could
induce a different composition of this layer,
determining different biomimetics surface properties.
This situation could occur, for example, when a
medical device is manipulated
in a bioenvironment whose composition
is externally controlled during a surgery \cite{Cloud2008}. In particular, we study 
the case in which the system is first equilibrated under
energy-depleted conditions and subsequently undergoes a sudden change
that reestablishes the normal conditions (Fig.~\ref{f6}).

At short times the energy-depleted system evolves until the  
equilibrium concentrations are reached. Under these conditions, as
discussed (Fig.~\ref{f2}), the dominant protein is IgG instead of Fib. At time $t_0$ 
we switch to normal conditions, forcing the system out of equilibrium.
As a consequence, the system re-enters a transitory situation in which
the concentrations $C_i^S$ evolve until they reach their new equilibrium values at long
times. In the specific case considered here, we observe a fast change
in the surface concentrations, with $C_F^S$ of Fib overcoming
$C_I^S$ of IgG, being the first, under normal conditions,
more stable on the surface than the second.
The final equilibrium concentrations are reached at large times.
We observe also a sudden change in $C_A^S$ of Alb, between the two
equilibrium concentrations characteristics of the two values of the
external parameters $T$. However, $C_A^S$ always equilibrates to a
value that is smaller then $C_I^S$ and $C_F^S$, consistent with
its long-time values in Fig.~\ref{f1}-\ref{f2}. 
By decreasing $X_P$, we find the
same qualitative behavior for a sudden energy-change, but  with 
the transient regime  extending to longer times, consistent
with Fig.~\ref{f2}. Hence, at experimental values of $X_P$ the
switching behavior would occur on time scales that are comparable to
those characteristic of the Vroman effect.

\subsection{Reversible vs Irreversible Adsorption}
 
We remark that our predictions about inverting the
Vroman effect by changing the experimental control parameters should hold
only if the protein adsorption on the surface is reversible. If the
adsorption is, instead, irreversible the change of external parameters
should not lead to a new composition of the protein layer. 
Indeed, under many practical conditions of interest for blood plasma, it would appear 
that the binding is indeed mostly irreversible \cite{Mono2011,Lundqvist2011}. 
Hence, 
the switching protocol proposed above represents
a possible experimental way to evaluate how strongly irreversible is the
adsorption process on a specific surface.

For an irreversible adsorption process, our findings predict that 
by appropriately controlling the parameters of the protein solution,
such as the amount of depleted energy, it is possible
to engineering a specific biomimetic covering of a surface.
Due to the irreversibility, 
the proteins, once adsorbed, cannot easily desorb from the
surface, even if the external conditions are modified. Therefore, 
it is feasible to cover a device surface with any desired protein composition,
targeted to a specific biomimetic property, by
selecting an appropriate initial condition. Subsequently,   
the device could be used under physiological conditions with no further
changes of the protein cover and its 
biomimetic properties. 

\section{Conclusions}

We study, by MD simulations of a coarse-grained model, the
Vroman effect for a ternary protein solution mixture, with 
Alb, IgG and Fib, in contact with a hydrophobic surface. We show that the
effect is the consequence of the differences among the proteins properties:
mass and size, affinity, diffusion constant, conformational changes.
These differences lead to a process of competitive adsorption on a
surface, in which the different 
families of proteins occupy sequentially the surface, replace each
other and diffuse at the surface,  until an equilibrium situation is reached.
By decreasing the total concentration of protein in the solution, keeping
the relative concentrations fixed, the time scales of the process
increase and the maxima of surface concentration for each family of
proteins occur at longer times.

Our model confirms the intuitive understanding that  the sequence of
surface occupation  is a consequence of a competition between
the smaller and faster, but less affine, proteins with the more
affine, but bigger and slower, proteins.  For
example, we test that by increasing the Alb affinity, or
artificially setting to the same value all the diffusion constants,
the Vroman effect disappears. Therefore, affinity and diffusion constant are
relevant protein parameters for the effect as can be deduced by
standard kinetics equation models.
 Nevertheless, our model reveals that the mechanisms of competitions
 are likely to be more complex that what the intuition would suggest,
when conformational changes occurs, with regimes during which small
and large proteins are not necessarily competing between them,
but are both competing with others in solution.

We find that 
the protein surface concentrations at equilibrium depend on 
external control parameters. 
In particular, we find that energy depletion induces a drastic change
in the composition of the covering protein-layer,  leading to an
inversion of the Vroman effect. Our results show that the inversion
can be used to quantify how strongly irreversible is the process of
surface adsorption of the proteins, an information useful in studies
of thromboembolic events \cite{Uniyal82}.
Furthermore, these results suggest
the possibility of engineering  
the composition of the protein layer covering a surface in a
controlled way, a feature particular relevant in biomimetic applications.

\section*{Acknowledgments}

We thank  C. $\AA$berg, F. Baldelli Bombelli, M. P. Monopoli and O. Vilanova
for discussions.
We acknowledge the support of EU FP7  grant NMP4-SL-2011-266737;
PV and GF of Spanish MEC grant FIS2012-31025  co-financed
FEDER.


\providecommand*{\mcitethebibliography}{\thebibliography}
\csname @ifundefined\endcsname{endmcitethebibliography}
{\let\endmcitethebibliography\endthebibliography}{}

\end{document}